\documentclass[twocolumn,tighten]{aastex61}
\usepackage{amssymb, amsmath,framed}

\usepackage{bm}
\expandafter\ifx\csname package@font\endcsname\relax\else
 \expandafter\expandafter
 \expandafter\usepackage
 \expandafter\expandafter
 \expandafter{\csname package@font\endcsname}
\fi
\def\bq{\begin{equation}}
\def\eq{\end{equation}}
\def\bqy{\begin{eqnarray}}
\def\eqy{\end{eqnarray}}

\def\calm{\mathcal{M}}
\def\caln{\mathcal{N}}

\begin{document}
\title{The Propitious Role of Solar Energetic Particles in the Origin of Life}

\correspondingauthor{Manasvi Lingam \& Chuanfei Dong}
\email{manasvi.lingam@cfa.harvard.edu \& dcfy@princeton.edu}

\author{Manasvi Lingam}
\affiliation{Institute for Theory and Computation, Harvard University, Cambridge MA 02138, USA}
\affiliation{Harvard-Smithsonian Center for Astrophysics, Cambridge, MA 02138, USA}

\author{Chuanfei Dong}
\affiliation{Department of Astrophysical Sciences, Princeton University, Princeton, NJ 08544, USA}
\affiliation{Princeton Center for Heliophysics, Princeton Plasma Physics Laboratory, Princeton University, Princeton, NJ 08544, USA}

\author{Xiaohua Fang}
\affiliation{Laboratory for Atmospheric and Space Physics, University of Colorado Boulder, Boulder, CO 80303, USA}

\author{Bruce M. Jakosky}
\affiliation{Laboratory for Atmospheric and Space Physics, University of Colorado Boulder, Boulder, CO 80303, USA}

\author{Abraham Loeb}
\affiliation{Institute for Theory and Computation, Harvard University, Cambridge MA 02138, USA}
\affiliation{Harvard-Smithsonian Center for Astrophysics, Cambridge, MA 02138, USA}

\begin{abstract}
We carry out 3-D numerical simulations to assess the penetration and bombardment effects of Solar Energetic Particles (SEPs), i.e. high-energy particle bursts during large flares and superflares, on ancient and current Mars. We demonstrate that the deposition of SEPs is non-uniform at the planetary surface, and that the corresponding energy flux is lower than other sources postulated to have influenced the origin of life. Nevertheless, SEPs may have been capable of facilitating the synthesis of a wide range of vital organic molecules (e.g. nucleobases and amino acids). Owing to the relatively high efficiency of these pathways, the overall yields might be comparable to (or even exceed) the values predicted for some conventional sources such as electrical discharges and exogenous delivery by meteorites. We also suggest that SEPs could have played a role in enabling the initiation of lightning. A notable corollary of our work is that SEPs may constitute an important mechanism for prebiotic synthesis on exoplanets around M-dwarfs, thereby mitigating the deficiency of biologically active ultraviolet radiation on these planets. Although there are several uncertainties associated with (exo)planetary environments and prebiotic chemical pathways, our study illustrates that SEPs represent a potentially important factor in understanding the origin of life.
\end{abstract}

\section{Introduction}
Our understanding of solar flares and associated phenomena, such as Solar Energetic Particles (SEPs) and Coronal Mass Ejections (CMEs), has improved greatly in recent times, both from a theoretical and observational standpoint \citep{WH12,Ream13,Pri14,CLHB,Benz17}. These discoveries have been supplemented by a wealth of data from the \emph{Kepler} mission, which surveyed $\sim 10^5$ stars and yielded detailed statistics concerning the frequency of large flares with energies $\gtrsim 10^{33}$ ergs (superflares) on M-, K-, and G-type stars \citep{Mae12,Shi13,CHMBS}.

Hence, there has been a concomitant increase in studies analyzing the effects of flares and superflares on planets situated in the habitable zone (HZ) of their host stars, i.e. the region in which liquid water can theoretically exist on the surface of the planet \citep{KWR93}. Since large flares are usually accompanied by bursts of high-energy radiation and particles, most studies have highlighted their deleterious effects from the perspective of habitability. The dangers posed by SEPs are especially significant since they may cause significant ozone depletion and biological damage \citep{Dart11,MT11,Ling2017}. However, recent evidence intriguingly suggests that flares could have played a beneficial role in the origin of life (abiogenesis) by delivering the requisite energy for the prebiotic synthesis of organic compounds \citep{BLM07,Aira16,NOSD,RWS17,Mana17,Gam17}.

It is therefore the goal of this paper to examine the potential role of SEPs, specifically energetic protons, in facilitating surface-based prebiotic chemistry. In this work, we shall not tackle non-surficial theories for the origin of life, such as hydrothermal vents \citep{MB08,RBB14}, since the SEP fluxes are not expected to be significant in these environments. We will focus on ancient and current Mars, primarily motivated by the fact that the habitability of both environments has been extensively investigated \citep{McK97,SFD08,WL13,Cock14}. Furthermore, some authors have argued that the biological potential of Noachian Mars was similar to, or slightly greater than, that of Hadean Earth \citep{JS98,NS01,GLE14,BK15} although its current value is several orders of magnitude smaller than present-day Earth \citep{SL02}; there also exists a remote possibility that life on Earth was seeded by Martian ejecta \citep{DM96,Mil00}. Subsequently, we will generalize our results to encompass exoplanets, with atmospheres resembling early Mars, orbiting low-mass stars (K- and M-dwarfs).

\section{Simulation setup and Results}

\begin{figure*}[!ht]
\centering
\includegraphics[scale=0.8]{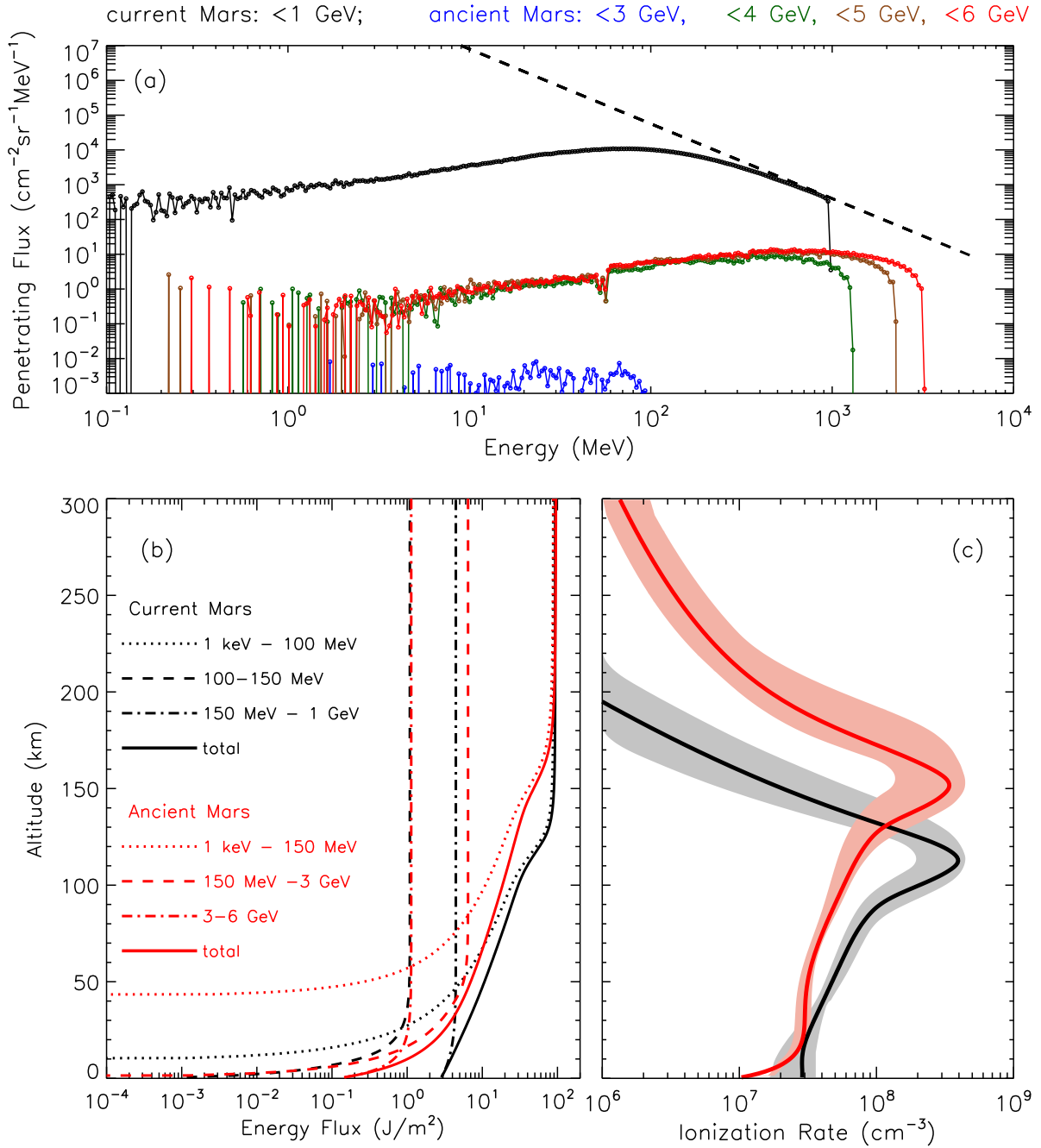}
\caption{Panel (a) illustrates the globally averaged differential number flux distribution at the planetary surface. Different colors represent cases with varying cutoffs marked at the top. The black dashed line shows the precipitating SEP energy spectral shape for reference. Panel (b) depicts the altitude profiles of penetrating SEP energy fluxes near the subsolar point under current (in black) and ancient (in red) atmospheric conditions. The energy fluxes are divided according to different energy ranges of incident particles at the topside boundary, as marked in the figure. Panel (c) presents the globally averaged ionization altitude profiles for $< 1$ GeV and $< 6$ GeV SEP precipitation in the current (black) and ancient (red) Martian atmosphere respectively. The shaded regions demarcate the range of values at different latitudes and longitudes.}
\label{FigFlux}
\end{figure*}

\begin{figure*}[!ht]
\centering
\includegraphics[scale=0.8]{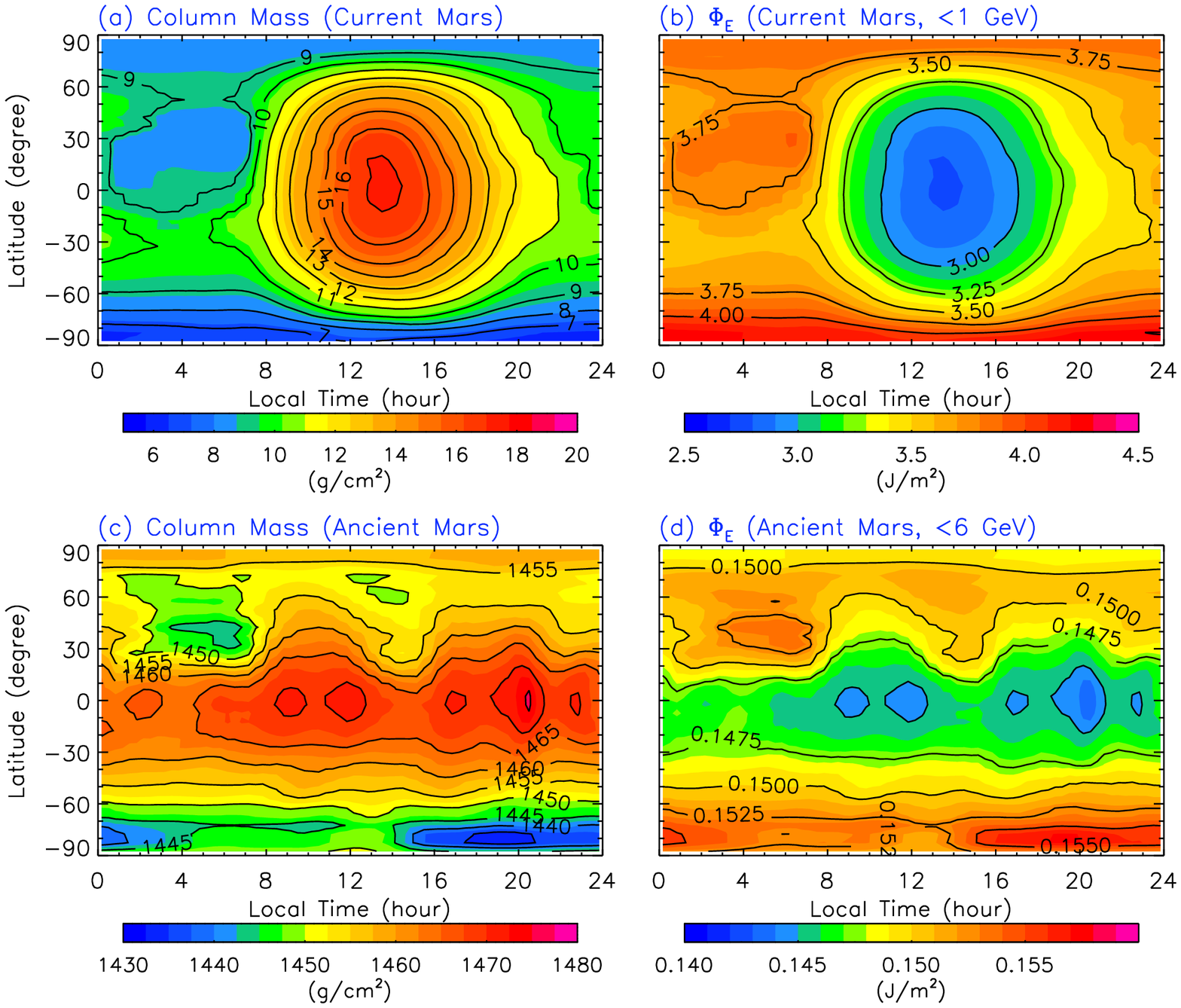}
\caption{Atmospheric column mass density and penetrating SEP flux at the surface for current ($< 1$ GeV) and ancient ($< 6$ GeV) Mars as a function of latitude and local time.}
\label{FigAsym}
\end{figure*}

We adopt the neutral atmosphere from the Mars Global Ionosphere Thermosphere Model (M-GITM) \citep{Bou15}, which is a 3-D whole atmosphere (ground-to-exobase) code that captures the Martian lower atmosphere and its thermosphere-ionosphere. It is particularly noteworthy that M-GITM, unlike previous General Circulation Models (GCMs), does not rely on the hydrostatic assumption and can therefore deal with large vertical velocities \citep{RDT06,DRRL}. This feature is particularly relevant since vertical winds must be taken into account when dealing with extreme space weather events, for e.g. CMEs and heating due to pickup ions \citep{FBJ13}.

To simulate the ``current'' Martian atmosphere, we use $P_s = 6$ mbar and $1$ EUV, while the ``ancient'' Mars atmosphere $\sim 4$ Gya is chosen to be $P_s \approx 1$ bar and $10$ EUV \citep{RGG05,Boe10}; here, $P_s$ denotes the surface pressure whereas $1$ EUV refers to the extreme ultraviolet (EUV) flux under current solar moderate conditions. It should be noted, however, that the exact value of $P_s$ for Noachian Mars is unknown, and it could have been lower than $1$ bar \citep{TKS}; a more tenous atmosphere would lead to an enhanced SEP flux at the surface for a given event. Next, we need to choose a particular SEP event to undertake our simulations. The input spectrum at the top of the atmosphere is adopted from the energetic January 2005 SEP event, and corresponds to an energy flux (per unit energy) of $5 \times 10^{11}$ pr cm$^{-2}$ MeV$^{-1}$ at $0.1$ MeV, while the associated spectral index is $-2.15$ \citep{Mew12}. We note that the 2005 SEP event has been studied previously as a representative example of a high-fluence solar proton event (associated with an X7 solar flare) in the context of prebiotic chemistry \citep{Aira16}. SEP events with total fluences greater than the January 2005 event have been documented \citep{Mew06}, and hence it might constitute a representative example for the active young Sun. However, it must be recognized that the SEP spectrum and initial flare energy are not necessarily correlated, since the causal mechanisms for SEPs are complex and varied \citep{Ream13}.

To calculate the penetration of SEPs through the atmosphere, we apply a continuously slowing down approximation (CSDA) energy loss model that was developed by \citet{JFS80} and recently improved by \citet{FLJ13}. The model has been extensively validated for energetic particle transport, showing excellent agreement with results from more sophisticated collision-by-collision calculations of the combined primary ion effects \citep{FLK,JDL17} and secondary electron effects \citep{LRA89}. Due to the lack of reliable collisional cross-section measurements, the collision-by-collision method cannot be extended to high-energy particles beyond $\sim 1$ MeV. In contrast, the stopping power data published by the National Institute of Standards and Technology covers a broader energy range, up to $10$ GeV \citep{Berg11}.

In Fig. \ref{FigFlux}(a), the penetrating SEP energy spectrum at the surface is shown for current and ancient Mars. From Fig. \ref{FigFlux}(b), it is seen that SEPs with initial energy lower than $\sim 150$ MeV (black dotted and dashed lines) are unable to penetrate through the current Martian atmosphere, and only those energetic particles with energies $\gtrsim 150$ MeV (black dashed-dotted line) can reach the surface; the same cutoff value has also been presented by the Mars Science Laboratory's \emph{Curiosity} rover group \citep{HZ14}. In contrast, the much thicker atmosphere assumed in the ancient epoch effectively prevents the penetration of $\lesssim 150$ MeV particles above $40$ km altitude (red dotted line). In fact, the threshold energy for incident SEPs being able to reach the surface through the ancient Martian atmosphere is elevated to $\sim 3$ GeV. Note that the threshold penetration energies of $\sim 150$ MeV and $\sim 3$ GeV, for current and ancient Mars respectively, vary slightly over the planet due to the spatial asymmetry of the atmospheres. The incident SEP spectrum at the top of the atmosphere is also shown for reference in Fig. \ref{FigFlux}(a).

In Fig. \ref{FigAsym}, column mass density and SEP energy deposition on the planetary surface as a function of latitude and longitude is presented. This is one of the notable features of our study since our simulations are fully 3-D and thus display signatures of asymmetry \citep{Bo06,Bou15}. For current and ancient Mars, the SEP energy flux broadly increases as one moves to higher latitudes. Moreover, for both cases, the energy flux is generally higher on the night-side relative to the day-side. The temperature difference between the day- and night-side affects the global neutral density distribution - see column mass density in Fig. \ref{FigAsym} - and consequently the SEP energy deposition at the planetary surface is altered, thus leading to the observed asymmetry. These asymmetric features are less pronounced for ancient Mars due to significant day-to-night transport caused by the strong day-side EUV heating; the transport leads to a relatively uniform distribution of the atmosphere around Mars. The resulting consequences are explored further in Sec. \ref{SSecOth}.

We have neglected the effects of weak planetary magnetic fields on SEPs in our simulations. Once the Martian dynamo stopped functioning $\sim 4.1$ Gya, the magnetic field strength is believed to have declined rapidly \citep{LFM08,FH11,Lil13}; note, however, that a later age has been proposed for the shutdown of the Martian dynamo by some authors \citep{SRM00}. The high energy of SEPs in conjunction with weak magnetic fields result in a very large gyroradius that may even exceed the planetary scale, suggesting that the deflection of SEPs by weak magnetic fields is minimal. Exoplanets around M-dwarfs could also be characterized by weak magnetic fields \citep{SBJ16}, and will be discussed later in Sec. \ref{SSecExo}.

\section{Implications for prebiotic chemistry}
We study some of the ensuing implications of the above results for Noachian Mars and other exoplanets (with similar atmospheres) in the HZ of their host stars.

\subsection{Energy source for prebiotic chemistry}\label{SSecEnS}
The origin of life required suitable energy sources for the synthesis of prebiotic compounds \citep{Ehr02,MS07,Lu16,Walk17}. Commonly studied energy pathways in this regard include solar radiation, shock heating from impacts, electrical discharges, radioactivity, volcanism and geochemical energy \citep{MU59,MS88,CS92,Pas12,RBD14}. It is therefore advantageous to estimate the energy available through SEPs that reach the surface and compare them against the aforementioned sources.

The energy flux $\Phi_E$ (in units of J m$^{-2}$ s$^{-1}$) can be estimated as follows:
\begin{equation} \label{EnFlux}
    \Phi_E = N \phi_E,
\end{equation}
where $\phi_E$ is the globally averaged energy per unit area deposited on the planetary surface (during a characteristic SEP event) and $N$ is the number of such SEP events per day.\footnote{In reality, $\Phi_E$ is time-averaged, and it is higher during the transient flaring period.} Our simulations yield $\phi_E = 1.5 \times 10^{-1}$ J/m$^2$ for ancient Mars, and we must now determine the value of $N$. The study of $\sim 10^5$ G-type stars by \emph{Kepler} yielded a power-law distribution for the occurrence frequency of superflares with a spectral index $\alpha$ between $-1.5$ and $-2.3$ \citep{Mae12,Mae15}. It has been suggested that superflares can occur on the Sun \citep{KS13,Mek15}, although the evidence remains disputed \citep{Sch12}. A similar power-law scaling has been inferred for regular flares on the Sun, but the spectral index is slightly different when compared to the corresponding estimate for superflares \citep{CAD93,Han11}.

As we are concerned with the young Sun, expected to have been more active $\sim 4.0$ Gya \citep{Gud07}, it seems reasonable to assume that it obeyed statistics similar to active G-type stars; the latter are predicted to have a superflare occurrence rate of $0.1$ per day \citep{Shi13}. However, we caution that the young Sun was not necessarily characterized by statistics similar to active solar-type stars studied by the \emph{Kepler} mission. With this caveat in mind, we use the above value to conclude that the frequency $N_0$ of Carrington-type events for the ancient young Sun may have been $N_0 \lesssim 40$ per day, which is lower by a factor of $\sim 6$ compared to \citet{Aira16}; these events are less powerful than superflares but occur more frequently with adequate energy release. Although large flares are often accompanied by CMEs and SEPs \citep{Em12,Ream13,DG16}, \emph{not} every such event will impact the planet. The number of impacting events is approximately estimated to be
\begin{equation} \label{Events}
    N = N_0 \sin^2\left(\frac{\theta}{2}\right),
\end{equation}
where the second factor is purely geometric, and is based on non-isotropic emission with an opening angle $\theta$. It stems from the solid angle fraction of the emitted particles, i.e. $\Omega/4\pi$ where we have $\Omega = \int_0^\theta \sin \theta' d\theta' \int_0^{2\pi} d\phi' = 2\pi \left(1-\cos \theta\right)$. The opening angle ranges between 20$^\circ$ and 120$^\circ$ and we select a fiducial value of $\theta \sim$ 47$^\circ$ based on observational evidence \citep{Yas04}. Upon substituting this value into (\ref{Events}), we find $N \lesssim 6$ events per day. In reality, $N$ will be even lower since other factors (e.g. magnetic field orientation) should be taken into consideration \citep{GYA07}. We will therefore normalize $N$ by $1$ event per day; see also \citet{OLHL} in this context. Using this data in (\ref{EnFlux}), we find
\begin{equation} \label{PhiMars}
    \Phi_E = 0.05\, f\,\left(\frac{N}{1\,\mathrm{day}^{-1}}\right)\,\mathrm{kJ}\,\mathrm{m}^{-2}\,\mathrm{yr}^{-1},
\end{equation}
where we have introduced the enhancement factor $f$ due to the following reason. We have implicitly assumed that only Carrington-type flares contribute to solar proton events impacting the planet. However, in reality, super-Carrington events with flare energies $\gtrsim 10^{33}$ ergs are expected to exist, as mentioned earlier \citep{KS13}. Although these events are rarer, they would be characterized by higher SEP fluences \citep{TMS16}. Thus, $f \gtrsim 1$ serves as the enhancement factor arising from these events, and is determined from the ratio of the cumulative and probability distribution functions describing the frequency of flares (with a given energy), and will depend on both $\alpha$ and the energy cutoff \citep{New05}. We will normalize $f$ by unity henceforth, since $f=1$ constitutes the lower bound.

The value of $\Phi_E$ in (\ref{PhiMars}) is lower than some energy fluxes on ancient Earth by 1-2 orders of magnitude \citep{DW10}, such as electrical discharges ($2.9\,\mathrm{kJ}\,\mathrm{m}^{-2}\,\mathrm{yr}^{-1}$) and volcanism ($5.4\,\mathrm{kJ}\,\mathrm{m}^{-2}\,\mathrm{yr}^{-1}$). However, it is worth noting that other energy sources are much higher, for e.g. the energy flux from solar UV radiation is about $7$ orders of magnitude greater. It must also be pointed out that the energy flux due to volcanism might have been much higher on early Mars \citep{HH14}. Here, we have compared $\Phi_E$ for Noachian Mars with other energy sources on Hadean-Archean Earth, since a comprehensive knowledge of the corresponding energy fluxes for ancient Mars is currently lacking. However, in light of the many unknowns, the above estimates should not be perceived as being definitive.

We can conjecture that  $\Phi_E^\mathrm{Earth} \gtrsim \Phi_E^\mathrm{Mars}$ because: (i) the SEP fluence falls off with the square of the distance \citep{FSWG,CCRW}, and (ii) the surface pressures and atmospheric compositions of ancient Mars and Earth appear to have been fairly similar \citep{McK10,AN12,Word16}. Since the SEP energy flux at the planetary surface depends on the incident SEP fluence and atmospheric properties, (i) and (ii) would therefore collectively imply that the SEP energy flux for Hadean-Archean Earth was higher than, or comparable to, that of Noachian Mars. Hence, the same conclusions discussed previously in the context of $\Phi_E^\mathrm{Mars}$ would also be valid for $\Phi_E^\mathrm{Earth}$, such as the comparison of the SEP energy flux against other sources (e.g. volcanism and lightning). 

It is possible to compute the number flux $\Phi_\caln$ of SEPs through a similar procedure. Thus, we estimate $\Phi_\caln$ via
\begin{equation} \label{Nflux}
    \Phi_\caln = f\, N\, \phi_\caln,
\end{equation}
where $f$ is the enhancement factor introduced earlier, $N$ is the number of Carrington-type SEP events per day, and $\phi_\caln \sim 7.5 \times 10^8$ cm$^{-2}$ from our simulations. Upon substituting these values into (\ref{Nflux}), we end up with
\begin{equation} \label{Nfval}
     \Phi_\caln = 8 \times 10^3\,f\,\left(\frac{N}{1\,\mathrm{day}^{-1}}\right)\,\mathrm{cm}^{-2}\,\mathrm{s}^{-1},
\end{equation}
and further implications are discussed in Sec. \ref{SSecOth}.

\subsection{Chemical pathways for prebiotic synthesis}\label{SSecChem}
Previously, we have argued that SEPs provide a valuable source of energy for prebiotic synthesis and that their energy flux is lower compared to lightning and volcanism. It is, however, important not only to evaluate the energy fluxes from different sources but also their efficiency in terms of chemical synthesis \citep{Deam97}.

SEPs are known to facilitate the formation of nitrogen oxides, for e.g. nitric oxide (NO) and nitrogen dioxide (NO$_2$), by reacting with atmospheric nitrogen \citep{CIR75,LPF05}.\footnote{Nitrous oxide (if produced) is a potent greenhouse gas \citep{CGF10,Aira16}, and may therefore partly provide a resolution of the ``faint young Sun'' paradox \citep{SM72,Kast10,Feu12}.} It was noted in \citet{Aira16} that these molecules reacted with the chemical species CH as follows:
\begin{eqnarray} \label{React}
\mathrm{NO + CH}\, &\rightarrow&\, \mathrm{HCN + O}, \nonumber \\
\mathrm{N_2O + CH}\, &\rightarrow&\, \mathrm{HCN + NO},
\end{eqnarray}
thereby resulting in the production of hydrogen cyanide (HCN). In addition, other pathways besides (\ref{React}) leading to SEP-driven HCN formation were also identified. HCN is important in prebiotic chemistry \citep{FH84} because it is required for the only known pathways leading to the prebiotic synthesis of (i) nucleic acids, (ii) proteins, and (iii) lipids through reductive homologation \citep{Pat15}; in turn, these molecules are postulated to have been important components of protocells. The chemical reactions leading to (i)-(iii) occur in the presence of UV radiation with hydrogen sulphide (H$_2$S) serving as the reductant. It has therefore been argued that HCN is an important ``feedstock'' molecule that may play a vital role in abiogenesis \citep{Suth16}. Hence, SEPs (and other energy sources) enable the production of HCN, which can undergo further UV-mediated chemical reactions to generate the above prebiotic compounds. However, a possible limitation is that HCN needs to be transported from the atmosphere to the surface, where the pathways of \citet{Pat15} are functional. 

Most of the basic ingredients presumably required for SEP-driven prebiotic synthesis, such as N$_2$, CO, CO$_2$, H$_2$O, H$_2$S and CH$_4$, are expected to have been present on Noachian Mars, although the exact composition remains very uncertain \citep{Owen92,JP01,FSJT,Form04,HZS07,WHA15,Word16}; these ingredients were also potentially available on ancient Earth for prebiotic synthesis \citep{ZSF10,AN12}. Moreover, the bioactive UV flux $\sim 4$ Gya emitted by the Sun is predicted to have been a few times times higher than its present-day value \citep{RSKS,RV16}, which can enable independent UV-mediated prebiotic chemistry \citep{RS17,Ran17}. Thus, it appears plausible that prebiotic pathways driven by SEPs (and UV radiation) were functional on ancient Mars and Earth, consequently rendering both planets conducive to the origin of life in this particular scenario.

In addition, gaseous mixtures - comprising of CO, CO$_2$, N$_2$ and H$_2$O - have also been shown in the laboratory (i.e., under controlled conditions) to yield a wide range of organic compounds when subjected to irradiation by energetic protons with energies greater than a few MeV. The organic molecules thus produced in the laboratory could, in principle, also be \emph{directly} synthesized by SEPs. A few examples of the salient organic compounds synthesized through this process include the following:
\begin{itemize}
    \item Uracil, guanine, adenine and cytosine \citep{KT90,Miy02}, which represent four of the five nucleobases of RNA and DNA. The absence of thymine (DNA only) might favor the emergence of RNA prior to DNA, in accordance with the RNA world hypothesis \citep{Joy02,Org04}. In contrast, meteorites ostensibly lack cytosine as well as thymine, thus making the eventual synthesis of both RNA and DNA (via this route) rather difficult \citep{PP16}.
    \item Amino acids such as alanine, aspartic acid, glycine and serine \citep{KKT98,Miy02}; most of these biomolecules are considered essential for protein synthesis \citep{WM81,ZZ08,HP09} and some authors have suggested that the co-evolution of amino acids and nucleic acids might have occurred \citep{KN09}. Some of these proteinogenic amino acids have also been produced by means of UV-mediated chemical pathways starting from HCN, as noted previously \citep{Mill13,Suth17}.
    \item Aromatic compounds, e.g. imidazole \citep{Kob95}, which could play an important role in abiogenesis \citep{ERCC}.
    \item Hydrogen cyanide, formaldehyde and complex amino acid precursors \citep{KKT98}.
\end{itemize}
Furthermore, if ammonia is unavailable in primordial planetary atmospheres, it has been suggested that irradiation by high-energy protons would be an important pathway for the synthesis of amino acid precursors \citep{KM01}. Lastly, the G-values associated with amino acid synthesis through irradiation by energetic protons ($\sim 0.01$) are amongst the highest documented for prebiotic pathways \citep{KT90}. This is an important point since it indicates that, despite the lower energy flux of SEPs, the resultant yield of organic molecules can be higher when compared to a few other energy sources. We will further quantify this statement by focusing on proteinogenic amino acids and nucleobases, i.e. the building blocks of proteins and nucleic acids respectively. 

However, before proceeding further, a few caveats are worth noting here. Given the significant uncertainties concerning the state of early Earth's atmosphere(s), the gaseous mixtures considered in the above papers could have been more reducing than that of the Hadean-Archean Earth. Second, the yields are dependent on the atmospheric composition, i.e. on the partial pressure of CO. Lastly, the experiments carried out relied on irradiation by protons of $\sim 3$ MeV, whereas the energies of SEPs can attain maximum values of a few GeV. As there is no experimental evidence available currently concerning prebiotic synthesis by particles at these energies, we will henceforth operate under the assumption that protons of different energies impact prebiotic chemistry uniformly; a similar assumption has also been invoked in analyzing endergonic chemical reactions driven by cosmic rays \citep{Miy02}. Although such an assumption will be utilized for carrying out quantitative estimates, we emphasize that there is a pressing need for follow-up experiments to address this important question.

The rate of amino acids synthesized, denoted by $\dot{\calm}_A$, is proportional to the deposited energy flux \citep{Kob95}. Using the G-values provided in \citet{KT90} and \citet{Miy02}, we obtain the following phenomenological relation
\begin{equation} \label{AmAcR}
    \dot{\calm}_A = 10^7\,\mathrm{kg}/\mathrm{yr}\,\left(\frac{\Phi_E}{0.1\,\mathrm{kJ}\,\mathrm{m}^{-2}\,\mathrm{yr}^{-1}}\right),
\end{equation}
for Earth-sized planets, where we have substituted the characteristic amino acid molar mass of $0.1$ kg/mol. The above formula follows from multiplying the energy flux with the efficiency of amino acid synthesis. This value is comparable to the yields of organic molecules from other energy sources as seen from Fig. 3 of \citet{Deam97}. However, it should be noted that (\ref{AmAcR}) deals \emph{solely} with the synthesis of amino acids, whereas the latter \citep{Deam97} depicted the estimates for all organic compounds. In a similar fashion, the amount of nucleobases produced can be evaluated accordingly. We end up with
\begin{equation} \label{NucR}
    \dot{\calm}_N = 10^4\,\mathrm{kg}/\mathrm{yr}\,\left(\frac{\Phi_E}{0.1\,\mathrm{kJ}\,\mathrm{m}^{-2}\,\mathrm{yr}^{-1}}\right),
\end{equation}
where $\dot{\calm}_N$ denotes the rate of nucleobases synthesized, and we have used the fact that the typical molar mass of nucleobases is $0.1$ kg/mol.

As a point of reference, let us consider the exogenous delivery of nucleobases and amino acids by meteorites. Using the delivery rate of intact organics from \citet{CS92} along with data from carbonaceous chondrites \citep{KL70,CP83,PS17}, we arrive at $\dot{\calm}_A \sim 300$ kg/yr and $\dot{\calm}_N \sim 2$ kg/yr \citep{Miy02}. Thus, provided that the constituent gases are available in sufficient amounts, prebiotic synthesis of amino acids and nucleobases by SEPs could have been approximately four orders of magnitude higher than delivery by meteorites; this follows from comparing the above values for meteorites with (\ref{AmAcR}) and (\ref{NucR}). 

For weakly reducing atmospheres we find that the yield of amino acids via electrical discharges is $\sim 5 \times 10^7$ kg/yr using the data from \citet{SM87} and \citet{Mill98}. A photochemical model was employed by \citet{TKZ11} to compute the surface deposition rates of HCN for different concentrations of methane, evaluated at different levels of carbon dioxide. Prior to the advent of methanogens, it was concluded that the prebiotic deposition rate of HCN was $\sim 10^{7}$ molecules cm$^{-2}$ s$^{-1}$. Upon converting this into kg/yr and using the conversion efficiency of HCN to amino acids \citep{SM87},  we find that the yield of amino acids would be $\sim 3 \times 10^7$ kg/yr. The values for electrical discharges and UV photochemistry are commensurate with the amount of amino acids that could be produced by means of SEP-mediated synthesis, as seen from (\ref{AmAcR}). In light of the many uncertainties surrounding prebiotic pathways and the composition of early Earth/Mars atmospheres, we emphasize that all of the preceding estimates should be regarded as being heuristic.

These organic compounds, especially nucleobases but also amino acids, will be deposited at very low concentrations on the planetary surface  (land and oceans) and subsequent prebiotic chemistry would face further challenges \citep{BS10}.\footnote{The putative role of aerosols in facilitating abiogenesis merits consideration \citep{Tuck,DTTV} in connection with the above limitation, since they function as non-equilibrium chemical reactors and can drive the concentration of reactants \citep{DETV,SAB13}.} In order for nucleotides to undergo rapid polymerization and yield nucleic acids, wet-dry cycles and thermal gradients are expected to play an important role on thermodynamic grounds via polymerase chain reactions \citep{KKLB,RD16}.\footnote{Alternatively, on a cold and wet ancient Mars \citep{Fai10,GMBH} or Earth \citep{ZSF10}, abiogenesis might have been facilitated through the concentration of prebiotic compounds by means of eutectic freezing \citep{MiCMi,Bad04,Pri07,MS08}. Ice and freeze-thaw cycles may have also played an important role in driving the assembly of RNA polymerase ribozymes \citep{TSB05,BBC,AWH13,MWH15}.} A wide range of putative environments endowed with operational cycles (and gradients) have been identified on Earth, for e.g. hydrothermal pools, intermountain valleys and beaches \citep{BC05,Adam,BKC12,DSMD,LiMa} to name a few. Most of these habitats also supply the requisite minerals (and nutrients) that can play an important role in abiogenesis by catalyzing polymerization \citep{FHLO,HS10,CSH12} and enabling homochirality \citep{HS03,Lamb08}.

However, it remains quite ambiguous as to whether analogous environments and mechanisms could have been prevalent on Noachian Mars \citep{CH10}. Surface water, minerals, and (perhaps) oceans are known to have existed intermittently on ancient Mars \citep{Bak01,SA05,TK09,DAH10,AHF13,EE14}. Conversely, geological explorations of Meridiani Planum by the Mars Exploration Rover \emph{Opportunity} suggest that the Martian paleoenvironment was likely to have been acidic, arid and oxidizing \citep{SK05,HM07,Ehl11}, while the salinity, ionic strength and chaotropic activity were potentially higher than the tolerance levels of current terrestrial organisms \citep{TKM08,BH15,FHCC}. These factors may have collectively posed difficulties for subsequent prebiotic chemical reactions and abiogenesis to occur on ancient Mars \citep{KC05}, although the environmental conditions at the Endeavour and Gale craters during the Noachian and/or Hesperian eras were possibly more favorable for the emergence of life \citep{Arv14,HG17}.

\subsection{Other implications of SEPs}\label{SSecOth}
Previously, we have outlined how SEPs may provide a direct energy source for prebiotic synthesis. However, there are other avenues by which they can indirectly contribute to the latter as well.

Physical mechanisms have been proposed wherein relativistic runaway electron avalanches (RREAs) associated with the formation of cosmic ray secondaries may lead to the initiation of lightning \citep{GMR92}. The basic principle behind RREAs is that runaway electrons can undergo M{\o}ller scattering and give rise to other free electrons which also exceed the runaway threshold, thereby giving rise to a cascade. However, in order to trigger RREAs, ``seed'' particles with sufficient energy are required. It has been hypothesized that these seed particles were primarily contributed by cosmic ray secondaries \citep{GZR99}. As SEPs can reach energies of a few GeV, it seems plausible that they could also serve as the seed particles for RREAs. There exists some recent observational evidence in this regard favoring positive correlations between elevated levels of SEPs (and solar activity) and lightning rates \citep{SSS,SHO14}. If this conjecture were valid, higher stellar activity would lead to enhanced lightning activity on (exo)planets \citep{HHA14}.

The relativistic runaway breakdown theory predicts that the flux of runaway electrons $\Phi_{RE}$ is linearly proportional to the flux of energetic seed particles $\Phi_\caln$ \citep{DU14}, and the latter is given by (\ref{Nfval}). The SEP number flux reaching the surface was probably comparable to the Galactic Cosmic Ray (GCR) flux \citep{San00}. If this were correct, solar proton events on ancient Mars (and Earth) may have contributed significantly to $\Phi_{RE}$ implying that SEPs, and therefore flares, could have played an important role in the initiation of lightning on these planets. The relevance of electrical discharges for prebiotic chemistry has been comprehensively studied \citep{Mil53,MU59,BC84,Bad13,Mill13}, especially in the context of nitrogen fixation; the latter's corresponding budget for ancient Mars is predicted to have been similar to that of early Earth \citep{SN05}.

From Fig. \ref{FigAsym}, the higher latitudes and night-side are characterized by slightly increased SEP energy fluxes at the surface for both current and ancient Mars. Since $\dot{\calm}_{A,N} \propto \phi_E$ follows upon combining (\ref{EnFlux}), (\ref{AmAcR}) and (\ref{NucR}), the synthesis of organic compounds could be higher at the above regions (up to a factor of $\sim 2$ for current Mars). Thus, the possibility that favorable conditions for abiogenesis existed at higher latitudes on ancient Mars, Earth and other exoplanets may merit consideration \citep{JNB03}, although it might be counterbalanced by the decline in reaction rates at relatively lower temperatures \citep{LiLo17}.

Hitherto, we have discussed the positive implications of SEPs, but we wish to reiterate that they can also be detrimental to life in surface environments. The production of hydrogen and nitrogen oxides by SEPs rapidly depletes the ozone layer \citep{Sol99,LPF05}, thereby enabling harmful UV-B and UV-C radiation to reach the surface and potentially triggering biological extinctions \citep{MT11,Ling2017}. SEPs could also trigger the formation of secondary energetic particles, and lead to high radiation doses, especially on planets in the HZ of low-mass stars, which can prove to be detrimental to certain complex lifeforms on the surface \citep{Atri,TSM17}. On the other hand, surficial organisms may adapt to such high-radiation environments by means of ultraviolet screening compounds \citep{CK99}. Furthermore, habitats in the deep biosphere or the oceans \citep{CM98} should be well-suited for protecting biota from ionizing radiation.

In addition, extreme space weather events also enhance atmospheric loss due to erosion by the solar/stellar wind \citep{DBM14,DBM15,DMB15,DHL17,DJL17,JGL15,MRF17}. In fact, if the atmosphere is altogether depleted in $\lesssim \mathcal{O}(100)$ Myr \citep{DLM17,LiLo17}, there might not be sufficient time for life to originate and evolve \citep{Ling7,Mana17}; see \citet{ST12} for a detailed Bayesian discussion of the constraints on the characteristic timescale for abiogenesis.

\subsection{Implications of SEPs for exoplanets}\label{SSecExo}
Let us now turn our attention to exoplanets in the HZ of other stars. We will focus primarily on M-dwarfs, as they are numerous in our Galaxy and detecting exoplanets around them is comparatively easier. 

Using the inverse-square scaling of SEP fluence \citep{FSWG} with orbital distance $a$, we have $\phi_{\caln,E} \propto a^{-2}$ and the available empirical evidence indicates that this approximation may be reasonably valid for exoplanets in the HZ of M-dwarfs \citep{YF17}. Assuming that the atmospheric properties are akin to that of ancient Mars, we obtain
\begin{equation} \label{PhiMD}
    \Phi_E = 50\, f\,\left(\frac{N}{1\,\mathrm{day}^{-1}}\right) \left(\frac{a}{0.05\,\mathrm{AU}}\right)^{-2}\,\mathrm{kJ}\,\mathrm{m}^{-2}\,\mathrm{yr}^{-1},
\end{equation}
and the last factor on the RHS accounts for the closer star-planet distance. We have retained the same normalization factor for $N$ given that M-dwarf exoplanets are impacted by $\lesssim 5$ events per day \citep{KOK16}. Similarly, we find that the number flux scales as
\begin{equation} \label{PhiNMD}
     \Phi_\caln = 7.3 \times 10^6\,f\,\left(\frac{N}{1\,\mathrm{day}^{-1}}\right) \left(\frac{a}{0.05\,\mathrm{AU}}\right)^{-2}\,\mathrm{cm}^{-2}\,\mathrm{s}^{-1}.
\end{equation}
From (\ref{PhiMD}) and (\ref{PhiNMD}), we see that the energy and number fluxes for exoplanets in the HZ of low-mass M-dwarfs are likely to be $\sim 100-1000$ times higher than that of ancient Mars (or Earth). In turn, this has a number of consequences delineated below.
\begin{itemize}
    \item The energy flux is expected to be higher than the corresponding values for most other sources on early Earth, with the exception of UV radiation. It can, for instance, exceed the energy fluxes for shock impacts and radioactivity.
    \item The number flux $\Phi_\caln$ is likely to be higher than that of GCRs by several orders of magnitude. In turn, SEPs could prove to be a fairly major player in the initiation of lightning.
    \item From (\ref{AmAcR}) and (\ref{NucR}), we see that $\dot{\calm}_{A,N} \propto \Phi_E$. Hence, SEPs might serve as the most significant mechanism for the synthesis of amino acids and RNA/DNA nucleobases on exoplanets around M-dwarfs; we find that $\sim 10^6-10^7$ kg/yr of nucleobases might be produced via SEPs and this value is comparable to, or slightly higher than, the upper bound of $\sim 10^6$ kg/yr estimated from Table 1 of \citet{NOSD}. 
    \item The above point is rendered even more important by the fact that UV-mediated prebiotic pathways are potentially inefficient on M-dwarf exoplanets, owing to the lower bioactive UV fluence (by a factor of $\sim 1000$) reaching the surface \citep{BLM07,RSKS}. It is, however, quite possible that flares may ameliorate this UV deficiency to some degree \citep{SSW,BLM07,RWS17}.
    \item Since $\Phi_E$ is proportional to $N$, as seen from (\ref{PhiMD}), it could imply that stars with higher flare activity are more conducive to prebiotic synthesis via SEPs (and even UV radiation). However, the same factors can also prove to be detrimental to the sustenance of complex biospheres \citep{Dart11}.
\end{itemize}
Most of the above conclusions concerning M-dwarf exoplanets are also likely to be applicable to planets in the HZ of K-dwarfs, although the respective estimates should be lowered by $1$-$2$ orders of magnitude. 

\section{Conclusions}
We carried out numerical simulations of ancient (and current) Mars to estimate the energy flux of SEPs (specifically protons) impacting the planetary surface and demonstrated that it is definitively lower than the contributions from other well-known energy sources like UV radiation and lightning. We also proved that energy deposition of SEPs on the surface displays a distinctive asymmetry with respect to both latitude and longitude.

We employed the energy flux of SEPs incident upon the surface of ancient Mars to arrive at certain noteworthy conclusions. By drawing upon detailed experimental evidence, we hypothesized that a wide range of prebiotic molecules, such as RNA/DNA nucleobases, aromatic compounds and amino acids, could be synthesized through irradiation by energetic protons. In particular, the estimated maximum yields of proteinogenic amino acids ($\sim 10^7$ kg/yr) and nucleobases ($\sim 10^4$ kg/yr) - the building blocks of proteins and RNA/DNA respectively - were potentially comparable to, or even orders of magnitude higher than, some of the widely explored pathways like electrical discharges and exogenous delivery by meteorites. We also suggested that SEPs may have played an important indirect role in prebiotic synthesis by initiating electrical discharges.

Subsequently, we generalized our analysis to encompass exoplanets with Mars-like atmospheres around K- and M-dwarfs. We conjectured that SEPs are perhaps an important factor in synthesizing organic compounds on exoplanets orbiting M-dwarfs, and that they might mitigate the paucity of bioactive UV radiation (and UV-mediated prebiotic chemistry). Hence, from the \emph{specific} viewpoint of prebiotic chemistry mediated by SEPs, it is tempting to conclude that these exoplanets are more conducive to abiogenesis.\footnote{Ironically, many of these characteristics are also possibly deleterious to the emergence and sustenance of complex surface-based biospheres on Mars \citep{DDW07,PV12} and Mars-like exoplanets in the HZ of M-dwarfs \citep{GS05,TVG16,TSM17}.}

However, in light of the many uncertainties and multiple interlinked factors involved concerning the origin of life, all of the above statements should be regarded with due caution. Detailed experimental and numerical follow-up studies based on the irradiation of gaseous mixtures resembling Noachian and M-dwarf (exo)planetary environments with MeV-GeV protons, in conjunction with remote sensing and \emph{in situ} searches for organic molecules on Mars \citep{BDMP,TK10,VW17},\footnote{In this context, it could be necessary to take into account the oxidizing agents (e.g. perchlorate ions) incorporated into the Martian regolith \citep{LN16}, ostensibly resulting in an oxidant extinction depth of $\lesssim 5$ m \citep{LL03}.} are necessary in order to fully assess the veracity of our results. Nevertheless, it seems plausible that SEPs might exert a profound influence (positive and/or negative) on the origin and evolution of life on habitable planets and moons \citep{HWK14}, both within and outside of our solar system.\\

\acknowledgments
The authors are very grateful to Stephen Bougher for generously providing access to the M-GITM model, and acknowledge valuable discussions with Andrew Knoll, Janet Luhmann, Michael Summers, Lei Dai, Takuya Shibayama and David Pawlowski. The detailed and insightful report furnished by the referee is also much appreciated. ML and AL were partly supported by grants from the Breakthrough Prize Foundation for the Starshot Initiative and Harvard University's Faculty of Arts and Sciences, and by the Institute for Theory and Computation (ITC) at Harvard University. CD was supported by the NASA Living With a Star Jack Eddy Postdoctoral Fellowship Program, administered by the University Corporation for Atmospheric Research. XF and BMJ acknowledge support from NASA's MAVEN mission. Resources for this work were provided by the NASA High-End Computing (HEC) Program through the NASA Advanced Supercomputing (NAS) Division at Ames Research Center.


\end{document}